\PassOptionsToPackage{table,dvipsnames}{xcolor}

\documentclass[sigconf]{acmart}
\usepackage{amsfonts}
\usepackage{subcaption}
\usepackage{enumitem}
\usepackage{stfloats}
\usepackage{bm}
\usepackage{makecell} 
\usepackage{multirow}
\usepackage{soul}
\sethlcolor{yellow}
\definecolor{yellow}{HTML}{f4c342}
\definecolor{GreenDark}{HTML}{187c19}
\definecolor{GreenMed}{HTML}{689f38}
\definecolor{GreenLight}{HTML}{8bc34a}
\definecolor{GreenSupLight}{HTML}{c5e1a5}

\AtBeginDocument{%
  }

\setcopyright{acmlicensed}
\copyrightyear{2025}
\acmYear{2025}
\setcopyright{acmlicensed}\acmConference[ICAIF '25]{6th ACM International Conference on AI in Finance}{November 15--18, 2025}{Singapore, Singapore}
\acmBooktitle{6th ACM International Conference on AI in Finance (ICAIF '25), November 15--18, 2025, Singapore, Singapore}
\acmDOI{10.1145/3768292.3770407}
\acmISBN{979-8-4007-2220-2/2025/11}

\begin{document}

\title{Time-Varying Factor-Augmented Models for Volatility Forecasting}


\author{Duo Zhang}
\email{dz2349@nyu.edu}
\orcid{0009-0005-5951-0851}
\affiliation{
  \institution{New York University}
  \city{New York}
  \state{New York}
  \country{USA}
}

\author{Jiayu Li}
\email{jl15681@stern.nyu.edu}
\orcid{0009-0002-6625-6167}
\affiliation{
  \institution{New York University}
  \city{New York}
  \state{New York}
  \country{USA}
}

\author{Junyi Mo}
\email{junyi.mo@stern.nyu.edu}
\orcid{0009-0001-2285-7019}
\affiliation{
  \institution{New York University}
  \city{New York}
  \state{New York}
  \country{USA}
}

\author{Elynn Chen}
\email{elynn.chen@stern.nyu.edu}
\orcid{0000-0002-7599-1828}
\authornote{Corresponding author}
\affiliation{
  \institution{New York University}
  \city{New York}
  \state{New York}
  \country{USA}
}

\renewcommand{\shortauthors}{Zhang, Li, Mo, Chen}

\begin{abstract}

Accurate volatility forecasts are vital in modern finance for risk management, portfolio allocation, and strategic decision-making. However, existing methods face key limitations. Fully multivariate models, while comprehensive, are computationally infeasible for realistic portfolios. Factor models, though efficient, primarily use static factor loadings, failing to capture evolving volatility co-movements when they are most critical. To address these limitations, we propose a novel, model-agnostic \emph{Factor-Augmented Volatility Forecast framework}. Our approach employs a time-varying factor model to extract a compact set of dynamic, cross-sectional factors from realized volatilities with minimal computational cost. These factors are then integrated into both statistical and AI-based forecasting models, enabling a unified system that jointly models asset-specific dynamics and evolving market-wide co-movements. Our framework demonstrates strong performance across two prominent asset classes—large-cap U.S. technology equities and major cryptocurrencies—over both short-term (1-day) and medium-term (7-day) horizons. Using a suite of linear and non-linear AI-driven models, we consistently observe substantial improvements in predictive accuracy and economic value. Notably, a practical pairs-trading strategy built on our forecasts delivers superior risk-adjusted returns and profitability, particularly under adverse market conditions.

\end{abstract}



\maketitle

\section{Introduction} 

From tick‑by‑tick hedging to long‑term investment strategy, reliable volatility forecasts are indispensable for guiding financial decisions. A vast literature—ranging from linear ARCH \cite{engle1982autoregressive}, GARCH \cite{bollerslev1986generalized}, and HAR \cite{corsi2009simple} to nonlinear neural networks such as LSTMs \cite{hochreiter1997long}—is typically applied on an asset-by-asset basis, modeling each asset in isolation. This univariate approach overlooks market-wide shocks that propagate rapidly across assets \cite{hamao1990correlations}, often biasing forecasts and understating systemic risk. Although multivariate extensions such as BEKK‑GARCH \cite{engle1995multivariate}, DCC‑GARCH \cite{engle2002dynamic}, Conditional Autoregressive Wishart \cite{golosnoy2012conditional}, 
and deep multivariate LSTM architectures capture shared dynamics, they are rarely used in real‑world settings due to parameter explosion, fragile estimation, and heavy tuning burdens \cite{bauwens2006multivariate,engle2019large}.

Factor models offer a natural remedy by efficiently summarizing co-movements through a low-dimensional set of latent drivers. Yet, two key limitations hinder their utility in volatility forecasting. First, most factor models are designed for returns, not volatility. Canonical linear frameworks such as Fama-French and Carhart \cite{fama1993common, carhart1997persistence}, and their modern machine‑learning successors \cite{gu2020empirical}, all distill factors from returns data. Consequently, the rich patterns of commonality in volatility—which often exhibit their own distinct dynamics—remain largely unexplored. Second, although a few recent pioneering studies do extend factor models to realized volatility \cite{atak2013factor, ding2025multiplicative}, they still impose static factor loadings, which fail to capture time-varying interdependencies—an especially critical flaw given that volatility co-movement evolves over time \cite{del2008dynamic}. Although recent econometric advances do provide time-varying loadings estimators \cite{mikkelsen2019consistent}, these techniques have so far been confined exclusively to asset returns. 

To bridge these gaps, we propose a model-agnostic \emph{Factor-Augmented Volatility Forecast framework} that captures evolving market-wide volatility shocks with computational tractability.
Unlike traditional factor approaches with static loadings, our method employs a time-varying factor model to extract latent factors directly from realized volatilities. This dynamic approach ensures that both the factors and their loadings adapt in real time to shifting market regimes. 
Compared to conventional multivariate techniques, our framework achieves this enhanced market awareness while preserving the parsimony of the base model, furthermore, introducing only a minimal number of extra parameters. The result is a robust, accurate, and interpretable framework that delivers computationally efficient volatility forecasts, suitable for practical deployment in dynamic financial environments. 

We evaluate the framework on two representative markets—U.S. technology equities and cryptocurrencies—for both 1-day and 7-day forecast horizons. Across all model-market-horizon combinations, factor augmentation consistently improves performance. Out-of-sample coefficient of determination ($R^2$) increases by up to 12.8\% for equities and 22.8\% for cryptocurrencies, with corresponding reductions in Mean Squared Error (MSE) and Quasi-likelihood loss (QLIKE). Furthermore, a volatility-scaled pairs-trading backtest confirms the framework's tangible economic value. Most strikingly, in a challenging market period, augmentation reversed an unprofitable strategy from an annualized loss of -5.5\% into a +7.3\% gain, flipping its Sharpe ratio from negative to positive. In brief, our key contributions are as follows: 
\begin{itemize}[noitemsep, topsep=0pt, parsep=0pt, partopsep=0pt]
    \item We introduce the first volatility forecasting framework that systematically integrates dynamic, cross-sectional information through factor augmentation. Our approach combines three unique innovations: (i) direct factor extraction from realized volatilities, distilling interpretable factors that capture volatility-specific commonalities; (ii) time-varying loadings that adapt to evolving market interdependencies; and (iii) a computationally efficient augmentation mechanism that ensures both robustness and scalability.
    \item Our framework achives superior statistical accuracy across all tested scenarios, and a practical pairs-trading backtest confirms that these accuracy gains translate directly into higher risk-adjusted returns, generating real-world economic value. 
    \item We provide a practical, end-to-end guide to factor extraction, selection, and integration, offering a clear roadmap for tailoring the framework to any forecasting model, horizon, or asset class.    

\end{itemize}

\section{Related Work}
Research in discrete-time volatility forecasting has evolved significantly. However, existing models present common trade-offs, which our framework is designed to address. 

\textbf{Univariate Models.} Frameworks from early AR models
to more sophisticated ARCH \cite{engle1982autoregressive}, GARCH \cite{bollerslev1986generalized, bollerslev1987conditionally}, and HAR \cite{corsi2009simple} have long been pillars of the field due to their interpretability. Extensions have incorporated features like mixed-frequency data \cite{conrad2020two, santos2014volatility} and long-memory processes \cite{baillie1996long}. However, by modeling each asset in isolation, these models suffer from a "univariate blind spot," ignoring the systemic co-movements that are crucial drivers of market-wide risk \cite{herskovic2016common, hamao1990correlations}. 

\textbf{Multivariate Models.} Multivariate models are designed to capture asset interplay directly. Within the GARCH family, the BEKK specification \cite{engle1995multivariate} guarantees positive-definite covariance matrices, while the Dynamic Conditional Correlation (DCC) model \cite{engle2002dynamic} offers a more feasible estimation process. AR-style alternatives like the Conditional Autoregressive Wishart (CAW) \cite{golosnoy2012conditional} provide further flexibility. While theoretically comprehensive, these models are often impractical for large systems due to the explosion of parameter count \cite{bauwens2006multivariate} and numerically unstable estimation \cite{engle2019large}.

\textbf{Machine Learning Models.} A parallel track has leveraged machine learning to capture the complex, nonlinear volatility patterns. Deep learning models like LSTM \cite{hochreiter1997long} have shown superior performance over classical models by modeling long-range dependencies in financial time series \cite{liu2019novel, mcnally2018predicting}. However, these models are not immune to the "curse of dimensionality." When applied in a fully multivariate setting, they face excessive parameterization, high overfitting risk, and prohibitive computational costs \cite{ge2023review}, inheriting the same fundamental limitations as their statistical counterparts.

\textbf{Factor Models.} 
Factor methods are the backbone of modern portfolio theory, distilling hundreds of correlated returns into a handful of common drivers and powering scalable asset allocation, hedging, and attribution. Widely used examples range from statistical factor‑GARCH variants such as OGARCH \cite{alexander2000primer} and GO‑GARCH \cite{van2002go} to recent factor‑structured deep networks \cite{feng2024deep}. While these models achieve parsimony by using latent factors, most of the existing work targets returns \cite{gu2020empirical,chen2020semiparametric,chen2019constrained,chen2022modeling,liu2022identification,chen2023statistical,chen2024semi,chen2024time,chen2024factor,kong2024teaformers}; factor models for volatility remain comparatively sparse and always assume time‑invariant loadings \cite{atak2013factor, ding2025multiplicative}. 

Our research directly confronts these trade-offs by proposing a factor augmentation framework. This approach enhances existing volatility models with a set of dynamic, cross-sectional factors, enabling the framework to capture systemic, time-varying market interdependence without adding additional computational overhead, thereby balancing predictive accuracy with practical feasibility.

\section{Data and Volatility Measurement}
\subsection{Data}
The empirical analysis is conducted on two high-frequency datasets, encompassing U.S. equities and major cryptocurrencies.

The equity dataset focuses on five prominent firms within the S\&P 500 Information Technology Sector, selected for their large market capitalization and historical sectoral comovements: Microsoft (MSFT), Advanced Micro Devices (AMD), Intel (INTC), Oracle (ORCL), and Cisco (CSCO). For these firms, we obtained high-frequency Consolidated Trade and Quote records from Wharton Research Data Service, covering the period from January 2018 to December 2020 at millisecond precision.

The cryptocurrency dataset, sourced from Kaiko, covers January 2018 to April 2021. This dataset comprises detailed, high-frequency trading data for five leading cryptocurrencies: Bitcoin (BTC), Ethereum (ETH), Ripple (XRP), Cardano (ADA), and Litecoin (LTC). Kaiko compiles transaction-level records from major digital asset exchanges, providing comprehensive data including timestamps, trade volumes, and direction.

\subsection{Volatility Measurement}
\subsubsection{Daily Realized Volatility}

Following \cite{andersen2001distribution}, we estimate daily realized volatility (RV) nonparametrically from high-frequency midpoint prices. After filtering out non-positive, crossed, and spurious quotes, midpoint prices for asset \(i\) on day \(t\) are sampled every five minutes:
\begin{equation}
\text{MID}_{i,t} = \frac{\text{BID}_{i,t} + \text{ASK}_{i,t}}{2}.
\end{equation}
The daily RV is then calculated as the square root of the sum of squared intraday returns: 
\begin{equation}
RV_{i,t} = \sqrt{\sum_{s=1}^{N} \left[\ln\left(\frac{\text{MID}_{i,t_s}}{\text{MID}_{i,t_{s-1}}}\right)\right]^{2}}\;,
\end{equation}
where \(\mathrm{MID}_{i,t_s}\) is the mid‐price at the \(s\)th five‐minute interval, \(N=78\) for a standard equity trading day and \(N=288\) for 24‐hour cryptocurrency markets. 

\subsubsection{Aggregated Realized Volatility}

To capture medium-term risk dynamics, we construct weekly realized volatility by averaging daily RVs over a 7-day window (\(h=7\)) following \citep{corsi2009simple,bollerslev2018risk}:
\vspace{-1pt}
\begin{equation}
RV_{t}^{7} = \frac{1}{7}\sum_{j=0}^{6} RV_{t-j}\;.
\end{equation}
\vspace{-1pt} 
This 7‑day moving average smooths daily noise, yielding a stable volatility gauge that matches common weekly rebalancing and reporting cycles.

\section{Methodology}
\subsection{Volatility Factor Analysis Across Assets}
\subsubsection{Time-Varying Factor Model}
Let \(\mathbf{y}_t = (RV_{1,t}, \dots, RV_{p,t})\) denote the \(p\)-dimensional vector of realized volatilities
at time \(t=1,\ldots,T\). Following the time‑varying matrix factor model of \cite{chen2024timevaryingmatrixfactormodels}, we posit the locally‑smooth factor representation: 
\begin{equation}
\label{eq:tvc_model}
\mathbf{y}_t \;=\; \mathbf{\Lambda}_t \mathbf{f}_t + \mathbf{\varepsilon}_t,
\end{equation}
where \(\mathbf{f}_t\in\mathbb{R}^k\) is the \(k\)-dimensional latent factors, and \(\mathbf{\varepsilon}_t\in\mathbb{R}^p\) is the idiosyncratic errors, assumed to be uncorrelated with \(\mathbf{f}_t\) but allowed to exhibit weak serial dependence. To accommodate a wide range of potential smooth temporal variation, we specify \(\mathbf{\Lambda}_t \in\mathbb{R}^{p\times k}\) to be the time‑varying loading matrix. Specifically, we model $\mathbf{\Lambda}_{t,i} =\mathbf{\Lambda}_{i\cdot}(t/T)$, where $\mathbf{\Lambda}_{i\cdot}(\cdot)$ is an unknown smooth function of $t/T$ on $[0,1]$ for each $i \in [p]$.

Let \(\mathbf{\Sigma}_f=\mathbb{E}(\mathbf{f}_t \mathbf{f}_t^{\top})\) and \(\mathbf{\Sigma}_\varepsilon=\mathbb{E}(\varepsilon_t\varepsilon_t^{\top})\). The second uncentered moment satisfies
\begin{equation}
\label{eq:second_moment}
\mathbf{\Sigma}_{y,t}
:= \mathbb{E}(\mathbf{y}_t\mathbf{y}_t^{\top})
= \mathbf{\Lambda}_t \mathbf{\Sigma}_f\mathbf{\Lambda}_t^{\top} + \mathbf{\Sigma}_\varepsilon .
\end{equation}
Under mild regularity conditions, the rank‑\(k\) signal component
\(\mathbf{\Lambda}_t\mathbf{\Sigma}_f\mathbf{\Lambda}_t^{\top}\) implies that the top \(k\) eigenvectors of
\(\mathbf{\Sigma}_{y,t}\) span the column space of \(\mathbf{\Lambda}_t\) up to rotation, as established in \cite{chen2024timevaryingmatrixfactormodels}.

In principle, one could estimate \(\mathbf{\Sigma}_{y,t}\) using kernel smoothing in \(t/T\) as in \cite{chen2024timevaryingmatrixfactormodels}.
For practical implementation, we instead adopt a \emph{rolling‑window} (uniform‑kernel)
approximation.
Fix a window length \(n\) and define \(W_t=\{t-n+1,\dots,t\}\).
We estimate the local covariance matrix by
\begin{equation*}
\widehat{\mathbf{\Sigma}}_{y,t}
= \frac{1}{n}\sum_{s\in W_t} \mathbf{y}_s\mathbf{y}_s^{\top}
= \sum_{s=1}^{T}K_{n}(s,t)\,\mathbf{y}_s\mathbf{y}_s^{\top},
\;
K_{n}(s,t)=\frac{1}{n}\,\mathbf{1}\{s\in W_t\}.
\end{equation*}
We extract the top-k eigenvectors \(\hat{\mathbf{u}}_{1,t},\ldots,\hat{\mathbf{u}}_{k,t}\) of
\(\widehat{\mathbf{\Sigma}}_{y,t}\). The estimated loading space and the contemporaneous factors  are then defined by
\begin{equation*}
\widehat{\mathbf{\Lambda}}_t \;=\; \sqrt{p}\,[\hat{\mathbf{u}}_{1,t},\ldots,\hat{\mathbf{u}}_{k,t}],
\qquad
\widehat{\mathbf{f}}_t \;=\; \frac{1}{p}\,\widehat{\mathbf{\Lambda}}_t^{\top}\mathbf{y}_t.
\end{equation*}
We apply this time-varying factor model separately to equities and cryptocurrencies, producing two distinct sets of $\widehat{\mathbf{\Lambda}}_t \text{ and } \widehat{\mathbf{f}}_t$.

\subsubsection{Divergent Factor Structures: Concentration in Crypto vs. Diffusion in Equity}
Both the equity and cryptocurrency markets exhibit hierarchical factor structures, yet they diverge significantly in risk concentration. This structural difference enables clear economic interpretations of the factors and provides a direct rationale for our horizon-dependent factor selection. 

\begin{figure*}[!t]
  \centering
  \begin{subfigure}[t]{0.24\textwidth}
    \includegraphics[width=\linewidth]{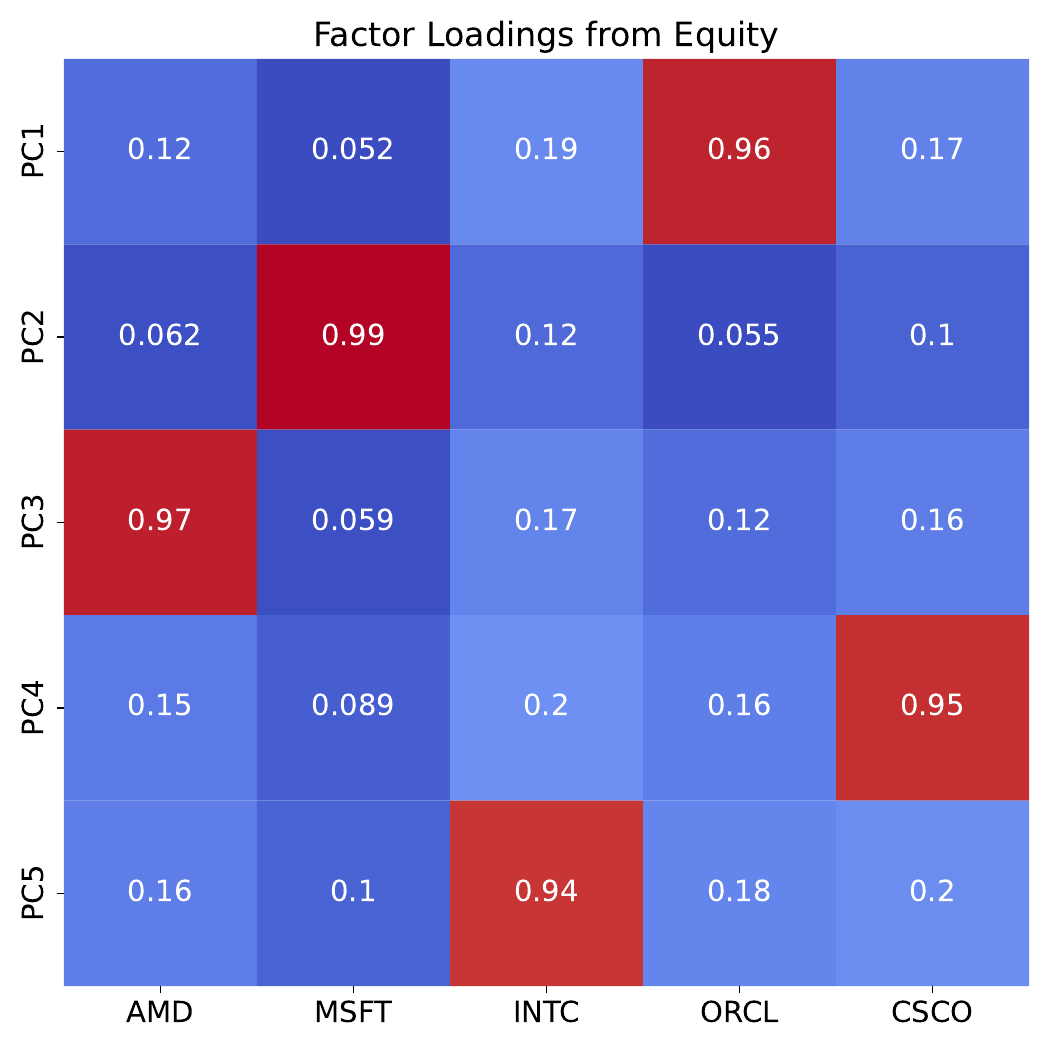}
    \caption{Factor Loadings for Equity}
    \label{fig: PC_Loading_Heatmap_Equity}
  \end{subfigure}\hfill
  \begin{subfigure}[t]{0.24\textwidth}
    \includegraphics[width=\linewidth]{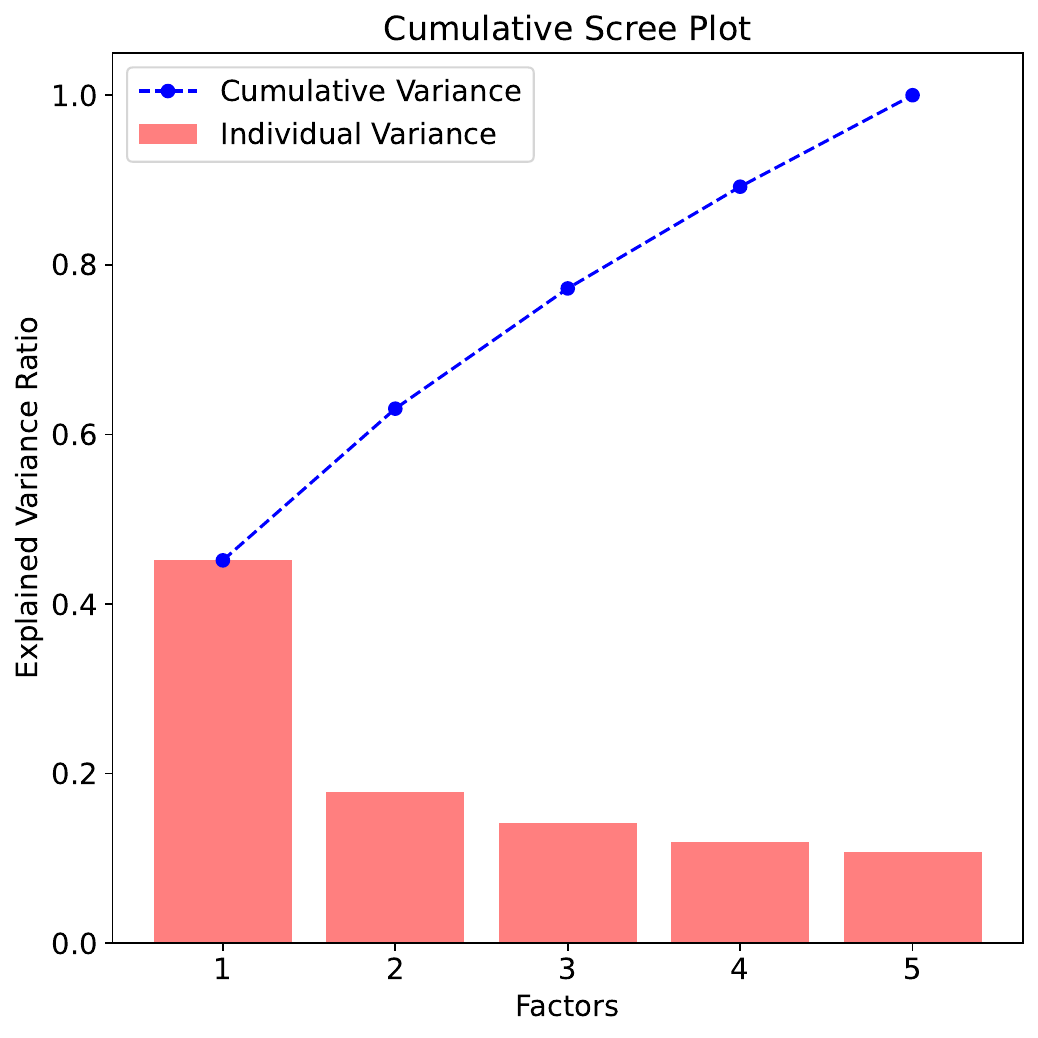}
    \caption{Explained Variance for Equity}
    \label{fig: Scree_Plot_Equity}
  \end{subfigure}\hfill
  \begin{subfigure}[t]{0.24\textwidth}
    \includegraphics[width=\linewidth]{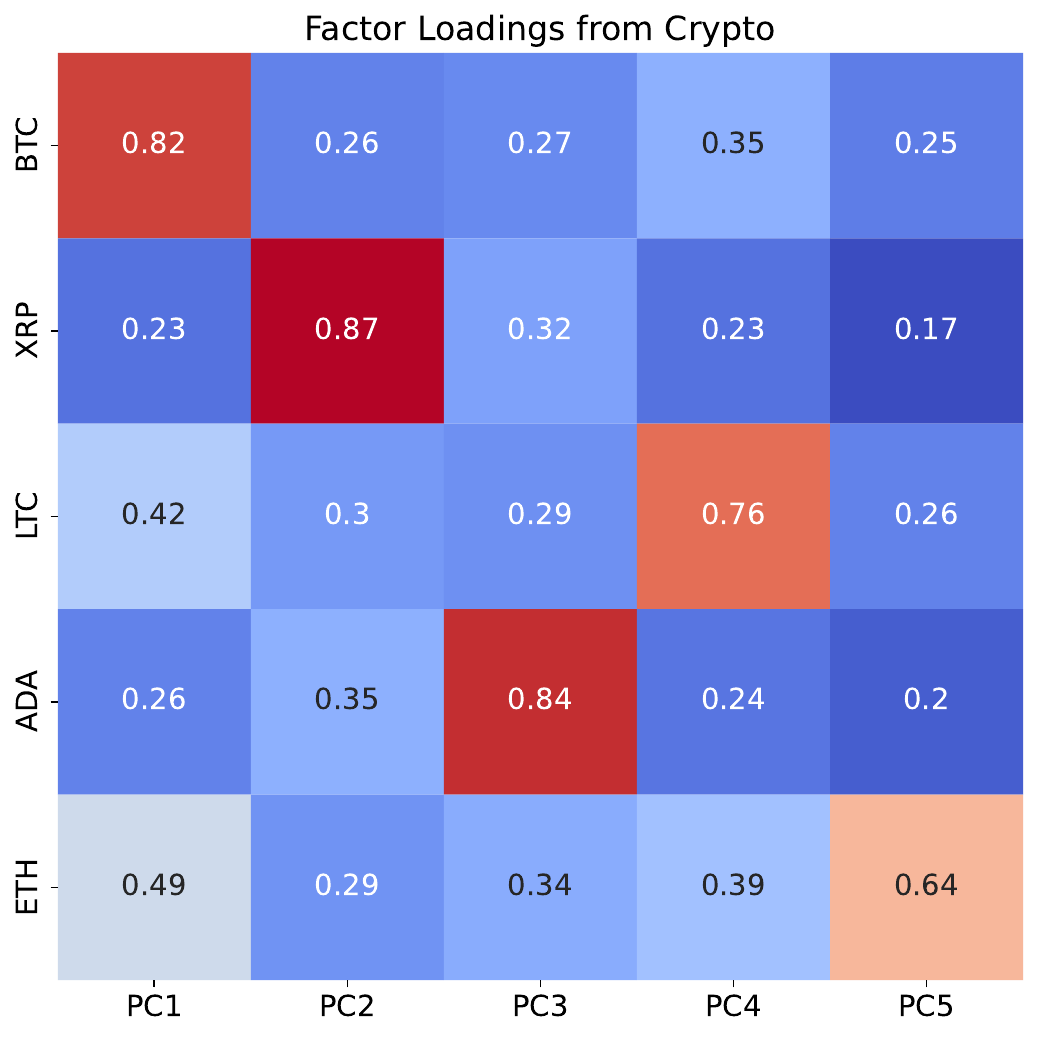}
    \caption{Factor Loadings for Crypto}
    \label{fig: PC_Loading_Heatmap_Crypto}
  \end{subfigure}\hfill
  \begin{subfigure}[t]{0.24\textwidth}
    \includegraphics[width=\linewidth]{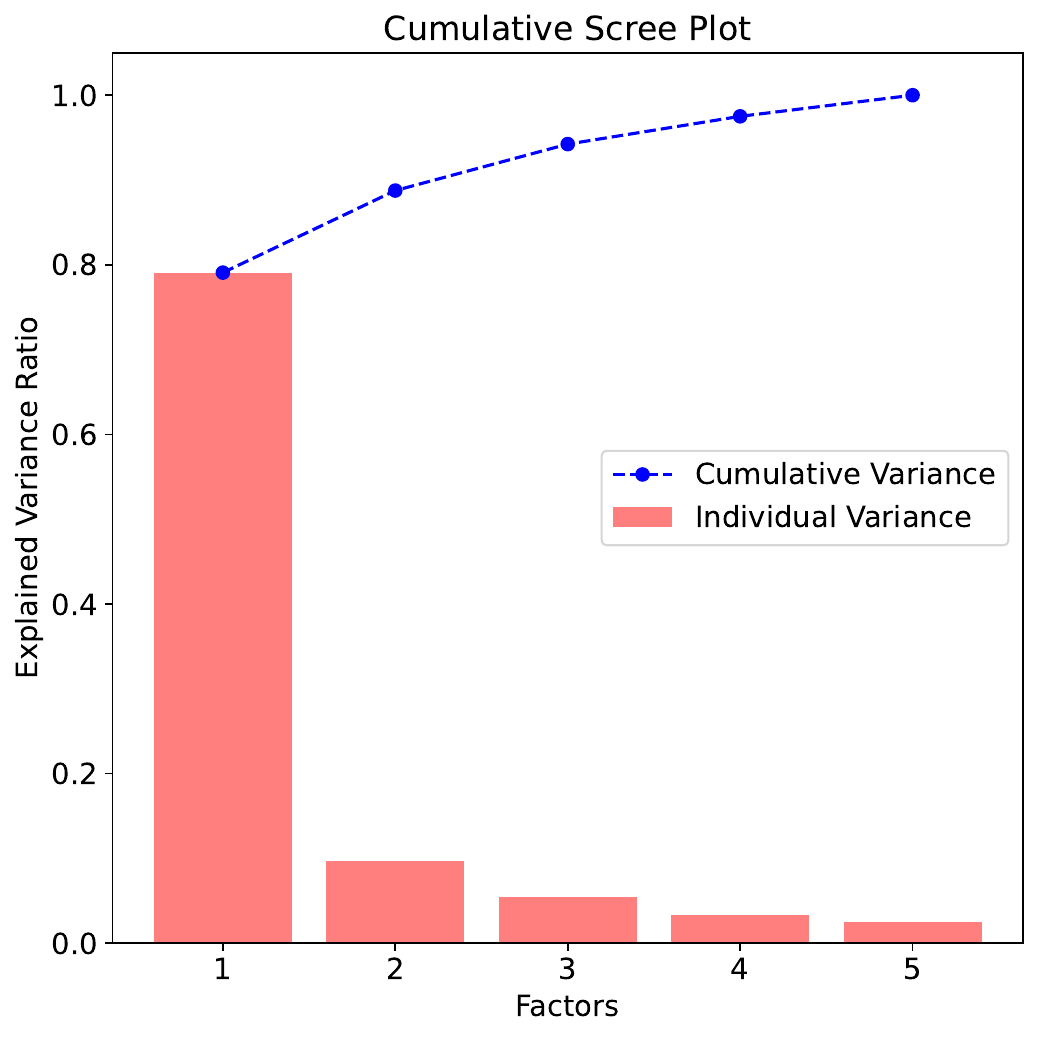}
    \caption{Explained Variance for Crypto}
    \label{fig: Scree_Plot_Crypto}
  \end{subfigure}
  \caption{Factor Loadings and Scree Plots for Equities and Crypto}
  \label{fig: PC}
\end{figure*}

The equity market's risk is diffuse. Its leading factor, \(\hat{\mathbf{f}}_{t}^{(1)}\), represents broad technology sector risk, explaining 45.2\% of the total variance, with firms like Oracle loading most heavily at 0.96. However, as the gradual decay in the scree plot Figure \ref{fig: Scree_Plot_Equity} illustrates, a single factor is insufficient. Subsequent components, from $\hat{\mathbf{f}}_{t}^{(2)} \text{ to }\hat{\mathbf{f}}_{t}^{(5)}$, capture firm-specific risks for companies like Microsoft and Intel, meaning a broader set of factors is needed to comprehensively describe market dynamics.

In stark contrast, cryptocurrency's risk is highly concentrated. A single, dominant systemic factor \(\hat{\mathbf{f}}_{t}^{(1)}\) serves as the primary market-wide driver, accounting for a commanding 
79.1\% of total variance. This factor is closely linked to the Bitcoin cycle, with its overwhelming influence confirmed by the sharp "elbow" in Figure \ref{fig: Scree_Plot_Crypto}. While subsequent factors isolate the dynamics of major altcoins like Ethereum and Ripple, their marginal contributions are more modest, highlighting a market driven by one principal force. 

These distinct factor structures provide a clear rationale for our horizon-dependent factor selection. For short-term (1-day) forecasts, we select only the single, dominant factor for each asset class to maximize the signal-to-noise ratio. For medium-term (7-day) forecasts, we employ an explained-variance threshold \cite{jolliffe2016principal, pukthuanthong2009global}. This involves selecting the minimum number of factors required to meet a predetermined cumulative variance target, which we tailor to each market's unique structure (e.g., 85\% for equities and 90\% for crypto). This data-driven approach ensures we capture significant secondary dynamics without overfitting on noise.

\subsection{Factor-Augmented Forecasting Models}

To demonstrate the flexibility and model-agnostic nature of our framework, we apply it uniformly to four benchmark models. The models forecast a common target variable,  $Y_{t+h}$, across two horizons $h\in\{1, 7\}$:  the next day's daily volatility $Y_{t+1} = RV_{t+1}$ and the aggregated weekly volatility for the upcoming week $Y_{t+7} = RV^7_{t+7}$.

The forecasts are based on two estimation approaches: an expanding window for the three statistical models and a fixed 80:20 train-test split for the more computationally intensive LSTM. 

\subsubsection{Autoregressive (AR)}
The AR model regress the target volatility $Y_{t+h}$ on the five most recent daily RVs to capture short-memory dynamics over a full trading week without over-parameterization: 
\begin{equation}
Y_{t+h} = \beta_0 + \sum_{i=0}^4 \beta_i RV_{t-i} + \epsilon_t.
\end{equation}
Given the AR model's simple linear structure, the factor-augmented version extends the baseline by directly incorporating the top-$S$ factors as additional predictors: 
\begin{equation}
    Y_{t+h} \;=\; \beta_0 + \sum_{i=0}^{4} \beta_i RV_{t-i}
            \;+\; \sum_{j\in [S]} \gamma_j \,\widehat{\mathbf{f}}_{j,t}
            \;+\; \epsilon_t.
\end{equation}
\noindent

\subsubsection{Heterogeneous Autoregressive (HAR)}

The HAR model captures multi-scale volatility persistence using daily, weekly, and monthly components \cite{corsi2009simple}:
\begin{equation}
Y_{t+h} = \beta_0 + \beta^d RV_t + \beta^w RV^{7}_t + \beta^m RV^{30}_t + \epsilon_t,
\end{equation}

\noindent
where $RV^d_t$, $RV^w_t$, and $RV^m_t$ represent daily, weekly, and monthly volatilities, respectively.

To align with the HAR’s multi-scale structure, our augmentation introduces factors constructed over corresponding daily and weekly horizons:
\begin{equation}
    \begin{split}
Y_{t+h} \;=&\; \beta_0 + \beta^{d}RV_t + \beta^{w}RV^{7}_t + \beta^{m}RV^{30}_t \\
          &\;+\; \sum_{j\in [{S}^{d}]} \gamma^{d}_{j}\,\widehat{\mathbf{f}}^{\,d}_{j,t}
            \;+\; \sum_{r\in [{S}^{w}]} \gamma^{w}_{r}\,\widehat{\mathbf{f}}^{\,w}_{r,t}
          \;+\; \epsilon_t,
\end{split}
\end{equation}

\noindent
where $\widehat{\mathbf{f}}^d_{j,t}$ and $\widehat{\mathbf{f}}^w_{r,t}$ denote the top-$S^d$ daily factors and the top-$S^w$ weekly factors, respectively.

\subsubsection{Mixed Data Sampling (MIDAS)}

The MIDAS model captures long-memory dynamics by incorporating many past observations through a parsimonious weighting scheme, specified as a Beta-polynomial distributed lag \cite{ghysels2007midas}:
\begin{equation}
Y_{t+h} = \beta_0 + \beta_1 \left[a(1)^{-1} a(L) RV_t \right] + \epsilon_t,
\end{equation}

\noindent
where lag weights $a(L)$ follow:
\begin{equation}
a_i = \left( \frac{i}{k} \right)^{\theta_1 - 1} \left( 1 - \frac{i}{k} \right)^{\theta_2 - 1} \frac{\Gamma(\theta_1 + \theta_2)}{\Gamma(\theta_1)\Gamma(\theta_2)}, \quad i = 1, \ldots, k,
\end{equation}

\noindent
with normalization $a(1) = \sum_{i=1}^k a_i$. $\theta_1 = 1$ and $\theta_2$ are optimized via in-sample grid search. Default $k = 30$ is extended with $k \in \{30, 50, 80\}$ for longer horizons.

To align with the MIDAS's core principle of parsimoniously weighting long-lag structures, we augment each factor through its own analogous Beta-polynomial $b(L)$:
\begin{equation}
    Y_{t+h} \;=\; \beta_0
               + \beta_1\!\bigl[a(1)^{-1} a(L) RV_t\bigr]
               + \sum_{j\in[S]} \gamma_j
                 \bigl[b(1)^{-1} b(L)\,\widehat{\mathbf{f}}_{j,t}\bigr]
               + \epsilon_t.
\end{equation}

\noindent

\subsubsection{Long Short-Term Memory (LSTM)}

Long Short‑Term Memory network is a gated recurrent architecture designed to capture long‑range dependencies in time‑series data. Each input vector concatenates a 7‑day window of daily realized volatility with the contemporaneous factor. This sequence passes through a stack of three LSTM layers, which learn the joint temporal patterns of volatility and factors. The final hidden state feeds a dense output layer that forecasts the target volatility $Y_{t+h}$.  

\subsection{Evaluation Metrics}  

Model performance is examined through three lenses:  
(i) statistical accuracy,  
(ii) statistical significance, and  
(iii) economic value.  

\subsubsection{Prediction Accuracy}  

Prediction accuracy is assessed with three widely used metrics: $R^2$, MSE, and QLIKE, which respectively capture explained variance, average prediction error, and distributional accuracy respectively \cite{patton2011volatility, zhang2024volatility}: 
\begin{align}
R^{2}   &= 1-\sum_{t}\left(\frac{RV^{h}_{t+h}-\widehat{RV}^{h}_{t+h}}
                     {RV^{h}_{t+h}-\overline{RV}^{h}_{t+h}}\right)^2,\\
\mathrm{MSE}   &= \frac{1}{T}\sum_{t}\left(RV^{h}_{t+h}-\widehat{RV}^{h}_{t+h}\right)^{2},\\
\mathrm{\mathrm{QLIKE}} &= \frac{1}{T}\sum_{t}\!\left[
                   \frac{RV^{h}_{t+h}}{\widehat{RV}^{h}_{t+h}}
                   -\ln\!\left(\frac{RV^{h}_{t+h}}{\widehat{RV}^{h}_{t+h}}\right)-1\right],
\end{align}

\subsubsection{Economic Implication}
We assess practical relevance using the out-of-sample Utility-of-Wealth ($\mathrm{UoW}$) metric:
\begin{equation}
\widehat{\mathrm{UoW}}=\frac{1}{T}\sum_{t=1}^{T}\!\left[
      \frac{SR^{2}}{\gamma}\frac{RV^{7}_{t+7}}{\widehat{RV}^{7}_{t+7}}
      -\frac{SR^{2}}{2\gamma}\!
       \left(\frac{RV^{7}_{t+7}}{\widehat{RV}^{7}_{t+7}}\right)^{2}\right],
\end{equation}
where we follow \cite{bollerslev2018risk} in setting the target Sharpe ratio to $\mathrm{SR}=0.4$, a typical long-run estimate for volatility-timing strategies, and choose risk aversion $\gamma=2$ to represent moderate investor preferences. Higher $\mathrm{UoW}$ denotes greater expected utility for a volatility-timing investor.

\subsubsection{Statistical Significance}
We assess systematic improvements with the Diebold–Mariano test\cite{diebold2002comparing}. For each horizon \(h\), we compute the one‐period loss differential: 
\begin{equation}
d_t \;=\;L\bigl(RV^h_{t+h},\widehat{RV}^h_{t+h,\mathrm{aug}}\bigr)
      \;-\;L\bigl(RV^h_{t+h},\widehat{RV}^h_{t+h,\mathrm{Baseline}}\bigr),    
\end{equation}
where
\begin{equation}
L(RV,\hat{RV})
=
\begin{cases}
(RV - \hat{RV})^2,
&\text{MSE‐based loss},\\[6pt]
\,-\left[\frac{SR^{2}}{\gamma}\frac{RV}{\hat{RV}}-\frac{SR^{2}}{2\gamma}\left(\frac{RV}{\hat{RV}}\right)^2\right],
&\text{Utility‐based loss}.
\end{cases}    
\end{equation}

\noindent The $\mathrm{DM}$ statistic: 
\begin{equation}
\mathrm{DM}
=\frac{\bar d}{\sqrt{\widehat{\mathrm{Var}}(\bar d)}},\quad
\bar d=\frac{1}{T}\sum_{t=1}^{T}d_t,    
\end{equation}
is asymptotically $\mathcal{N}(0,1)$ under $H_0:E[d_t]=0$. A positive DM value indicates that the factor-augmented forecasts significantly outperform the unaugmented benchmarks.


\section{Experiment Result}

\begin{table*}[htbp]
\centering
\small                            
\setlength{\tabcolsep}{4pt}       
\renewcommand{\arraystretch}{0.9} 
\caption{Cryptocurrency Volatility Forecast Accuracy: 1- and 7-Day Out-of-Sample Results}
\label{tab:Crypto_1d7d}
\resizebox{\linewidth}{!}{%
  \begin{tabular}{l | *{15}{c}}
        \midrule
    \multicolumn{16}{c}{\textbf{Panel A: Crypto in 1‐Day Horizon}} \\
    \midrule
      & \multicolumn{3}{c}{\textbf{BTC}} 
      & \multicolumn{3}{c}{\textbf{ETH}} 
      & \multicolumn{3}{c}{\textbf{XRP}} 
      & \multicolumn{3}{c}{\textbf{ADA}} 
      & \multicolumn{3}{c}{\textbf{LTC}} \\
    \cmidrule(lr){2-4} \cmidrule(lr){5-7} \cmidrule(lr){8-10} 
    \cmidrule(lr){11-13} \cmidrule(lr){14-16}
      & $\bm{R^2}$ & \textbf{MSE}   & \textbf{QLIKE}  
      & $\bm{R^2}$ & \textbf{MSE}   & \textbf{QLIKE}  
      & $\bm{R^2}$ & \textbf{MSE}   & \textbf{QLIKE}  
      & $\bm{R^2}$ & \textbf{MSE}   & \textbf{QLIKE}  
      & $\bm{R^2}$ & \textbf{MSE}   & \textbf{QLIKE}    \\
    \specialrule{\lightrulewidth}{1pt}{0.5pt}
    \textbf{AR}       &  43.60 & 3.27 & 0.0802 & 40.34 & 3.50 & 0.0687 & 45.06 & 7.69 & 0.0842  &   41.33 &  4.16  &  0.0499  & 39.95 & 4.13 & 0.0658\\ 
    \textbf{AR‐Aug}   &  \cellcolor{GreenLight}47.57 & \cellcolor{GreenLight}2.99 & \cellcolor{GreenSupLight}0.0770 & \cellcolor{GreenSupLight}42.34 & \cellcolor{GreenSupLight}3.38 & \cellcolor{GreenSupLight}0.0679 & \cellcolor{GreenLight}48.58 & \cellcolor{GreenLight}7.15 & \cellcolor{GreenLight}0.0789 & \cellcolor{GreenSupLight}42.01 & \cellcolor{GreenSupLight}4.13 &  \cellcolor{GreenSupLight}0.0498  &  \cellcolor{GreenSupLight}40.02 & \cellcolor{GreenSupLight}4.12 &  \cellcolor{GreenSupLight}0.0654  \\
    \specialrule{\lightrulewidth}{0.5pt}{0.5pt}
    \textbf{HAR}      &  39.52 & 3.12  & 0.0831 & 35.93 &  3.41  & 0.0691  & 47.58  & 6.98  & 0.0779  &  40.51 &  4.16 &  0.0497  & 34.08 & 4.09  & 0.0640  \\
    \textbf{HAR‐Aug}  & \cellcolor{GreenMed}43.93  & \cellcolor{GreenLight}0.88 &  \cellcolor{GreenSupLight}0.0797 &  \cellcolor{GreenMed}40.43 & 	\cellcolor{GreenLight}3.19 & \cellcolor{GreenSupLight}0.0681 & \cellcolor{GreenSupLight}49.73 & \cellcolor{GreenLight}6.63 & \cellcolor{GreenSupLight}0.0778 & \cellcolor{GreenSupLight}41.09 & \cellcolor{GreenSupLight}4.14 & \cellcolor{GreenSupLight}0.0497 & \cellcolor{GreenLight}37.41 & \cellcolor{GreenLight}3.85 & \cellcolor{GreenSupLight}0.0640 \\
    \specialrule{\lightrulewidth}{0.5pt}{0.5pt}
    \textbf{MIDAS(k=30)}   &  42.20 & 2.96 & 0.0828 & 39.07 & 3.29 &  0.0691 & 45.95 & 7.21 & 0.0843 & 40.94 & 4.19 & 0.0508 & 33.58 & 4.03 & 0.0683  \\
    \textbf{MIDAS‐Aug(k=30)}    &  \cellcolor{GreenSupLight}42.32 & \cellcolor{GreenSupLight}2.96 & \cellcolor{GreenSupLight}0.0827 & \cellcolor{yellow}38.93 & \cellcolor{GreenSupLight}3.30 & \cellcolor{yellow}0.0693 & \cellcolor{GreenSupLight}46.00 & \cellcolor{GreenSupLight}7.21 & \cellcolor{yellow}0.0844 & \cellcolor{GreenSupLight}40.94 & \cellcolor{GreenSupLight}4.19 & \cellcolor{GreenSupLight}0.0508 & \cellcolor{GreenSupLight}34.12 & \cellcolor{GreenSupLight}4.01 & \cellcolor{GreenSupLight}0.0679 \\
    \specialrule{\lightrulewidth}{0.5pt}{0.5pt}
    \textbf{LSTM}     &  53.88 & 2.13 & 0.0603 &  51.24 & 2.75 & 0.0551 & 50.73 & 13.20 & 0.0824 &  45.12 & 6.12 & 0.0550 & 51.64 & 3.66 & 0.0530 \\
    \textbf{LSTM‐Aug} & \cellcolor{GreenLight}57.70 & \cellcolor{GreenLight}1.99 & \cellcolor{GreenLight}0.0564 & \cellcolor{GreenLight}55.72 & \cellcolor{GreenSupLight}2.64 & \cellcolor{GreenLight}0.0506 &  \cellcolor{GreenLight}53.57 & \cellcolor{GreenSupLight}12.66 & \cellcolor{GreenSupLight}0.0800 & \cellcolor{GreenSupLight}47.34 & \cellcolor{GreenSupLight}6.08 & \cellcolor{GreenSupLight}0.0529 & \cellcolor{GreenSupLight}53.64 & \cellcolor{GreenSupLight}3.65 & \cellcolor{GreenSupLight}0.0519 \\
    \specialrule{\lightrulewidth}{0.5pt}{1.5pt}
    \multicolumn{16}{c}{\textbf{Panel B: Crypto in 7‐Day Horizon}} \\
    \specialrule{\lightrulewidth}{1pt}{0.5pt}
    \textbf{AR}        & 48.63 & 1.87 & 0.0677 & 39.98 & 2.08 & 0.0552 & 41.78 & 5.53 & 0.0801 & 38.37 & 2.89 & 0.0426 & 39.21 & 2.24 & 0.0475 \\
    \textbf{AR‐Aug}    &  \cellcolor{GreenSupLight}50.89 & \cellcolor{GreenSupLight}1.78 & \cellcolor{GreenSupLight}0.0648 & \cellcolor{GreenSupLight}41.85 & \cellcolor{GreenSupLight}2.03 & \cellcolor{GreenSupLight}0.0544 & \cellcolor{GreenSupLight}43.07 & \cellcolor{GreenSupLight}5.49 & \cellcolor{GreenSupLight}0.0792 & \cellcolor{GreenLight}40.70 & \cellcolor{GreenSupLight}2.78 & \cellcolor{GreenSupLight}0.0414 & \cellcolor{GreenLight}42.38 & \cellcolor{GreenSupLight}2.14 & \cellcolor{GreenSupLight}0.0453  \\
    \specialrule{\lightrulewidth}{0.5pt}{0.5pt}
    \textbf{HAR}      & 44.96 & 2.00 & 0.0793  & 38.25 & 2.15 & 0.0571 & 45.63 & 5.24 & 0.0782 & 36.71 & 2.96 & 0.0442 & 39.14 & 2.26 & 0.0492 \\
    \textbf{HAR‐Aug}   & \cellcolor{GreenLight}47.38 & \cellcolor{GreenSupLight}1.94 & \cellcolor{GreenLight}0.0737 & \cellcolor{yellow}37.90 & \cellcolor{yellow}2.17 & \cellcolor{yellow}0.0577 & \cellcolor{GreenSupLight}46.58 & \cellcolor{yellow}5.29 & \cellcolor{GreenSupLight}0.0782 & \cellcolor{GreenSupLight}38.03 & \cellcolor{GreenSupLight}2.90 & \cellcolor{GreenSupLight}0.0435 & \cellcolor{GreenSupLight}40.36 & \cellcolor{GreenSupLight}2.22 & \cellcolor{GreenSupLight}0.0474  \\
    \specialrule{\lightrulewidth}{0.5pt}{0.5pt}
    \textbf{MIDAS(k=30)}   &  49.40 & 1.84 & 0.0670 & 40.94 & 2.06 & 0.0546 & 43.26 & 5.47 & 0.0796 & 38.92 & 2.86 & 0.0424 & 40.82 & 2.20 & 0.0469 \\
    \textbf{MIDAS‐Aug(k=30)}    &  \cellcolor{GreenLight}53.26 & \cellcolor{GreenLight}1.70 & \cellcolor{GreenLight}0.0634 & \cellcolor{GreenLight}44.04 & \cellcolor{GreenLight}1.95 & \cellcolor{GreenLight}0.0516 & \cellcolor{GreenLight}47.23 & \cellcolor{GreenLight}5.09 & \cellcolor{GreenLight}0.0724 & \cellcolor{GreenDark}47.01 & \cellcolor{GreenMed}2.47 & \cellcolor{GreenLight}0.0390 & \cellcolor{GreenMed}45.08 & \cellcolor{GreenLight}2.07 & \cellcolor{GreenLight}0.0438  \\
    \specialrule{\lightrulewidth}{0.5pt}{0.5pt}
    \textbf{MIDAS(k=50)}   &  49.64 & 1.83 & 0.0670 & 40.90 & 2.06 & 0.0546 & 43.26 & 5.52 & 0.0804 & 39.00 & 2.86 & 0.0423  & 40.69 & 2.20 & 0.0470 \\
    \textbf{MIDAS‐Aug(k=50)}    & \cellcolor{GreenLight}52.95 & \cellcolor{GreenLight}1.71 & \cellcolor{GreenLight}0.0621  & \cellcolor{GreenLight}43.80 & \cellcolor{GreenLight}1.95 & \cellcolor{GreenLight}0.0514 & \cellcolor{GreenLight}47.34 & \cellcolor{GreenLight}5.05 & \cellcolor{GreenLight}0.0727 & \cellcolor{GreenDark}47.29 & \cellcolor{GreenMed}2.47 & \cellcolor{GreenLight}0.0389 & \cellcolor{GreenLight}44.30 & \cellcolor{GreenLight}2.07 & \cellcolor{GreenLight}0.0443  \\
    \specialrule{\lightrulewidth}{0.5pt}{0.5pt}
    \textbf{MIDAS(k=80)}   & 49.58 & 1.83 & 0.0673  & 40.96 & 1.95 & 0.0516 & 42.73 & 5.09 & 0.0724  & 38.92 & 2.48 & 0.0390 & 40.80 & 2.04 & 0.0438  \\
    \textbf{MIDAS‐Aug(k=80)}    &  \cellcolor{GreenLight}52.95 & \cellcolor{GreenLight}1.71 & \cellcolor{GreenLight}0.0628 &  \cellcolor{GreenLight}43.72 & \cellcolor{GreenSupLight}1.96 & \cellcolor{GreenSupLight}0.0518 & \cellcolor{GreenMed}47.33 & \cellcolor{GreenLight}5.08 & \cellcolor{GreenMed}0.0736  & \cellcolor{GreenDark}47.79 & \cellcolor{GreenMed}2.45 &  \cellcolor{GreenLight}0.0384 & \cellcolor{GreenLight}44.31 & \cellcolor{GreenLight}2.07 & \cellcolor{GreenSupLight}0.0447 \\
    \specialrule{\lightrulewidth}{0.5pt}{0.5pt}
    \textbf{LSTM}     &  59.31 & 1.37 & 0.0419 & 46.25 & 2.28 & 0.0441 & 42.81 & 1.43 & 0.1222 & 37.32 & 6.28 & 0.0629  & 52.39 & 2.38 & 0.0401   \\
    \textbf{LSTM‐Aug} & \cellcolor{GreenSupLight}61.66 & \cellcolor{GreenLight}1.30 & \cellcolor{GreenMed}0.0374 & \cellcolor{GreenMed}51.89 & \cellcolor{GreenMed}2.03 & \cellcolor{GreenMed}0.0385 & \cellcolor{GreenSupLight}44.25 & \cellcolor{GreenSupLight}1.41 & \cellcolor{GreenSupLight}0.1186 & \cellcolor{GreenMed}41.16 & \cellcolor{GreenSupLight}6.19 & \cellcolor{GreenSupLight}0.0616 & \cellcolor{GreenMed}59.83 & \cellcolor{GreenDark}2.38 & \cellcolor{GreenDark}0.0340 \\
    \specialrule{\lightrulewidth}{0pt}{0pt}
  \end{tabular}%
}
\begin{flushleft}
  \footnotesize\emph{Note:} All MSE values are reported in units of $10^{-4}$.\quad Improvement: \textcolor{GreenDark}{\rule{1.2em}{1.2em}}\ $\geq +15\%$ \quad
    \textcolor{GreenMed}{\rule{1.2em}{1.2em}}\ $\geq +10\%$\quad
    \textcolor{GreenLight}{\rule{1.2em}{1.2em}}\  $\geq +5\%$ \quad \textcolor{GreenSupLight}{\rule{1.2em}{1.2em}}\  $>\ 0\%$ \quad \textcolor{yellow}{\rule{1.2em}{1.2em}}\ $\leq -5\%$ 
\end{flushleft}
\end{table*} 

\subsection{A Universal Performance Boost Across All Models and Horizons}

Factor augmentation universally and significantly boosts forecast accuracy across every model, asset class, and horizon tested, proving its effectiveness for both short- and medium-term predictions, as shown in Tables~\ref{tab:Crypto_1d7d} and \ref{tab:Equity_1d7d}.

Specifically, in 1‑day forecasts, the benefits of factor augmentation are most pronounced for short-memory models, with HAR and LSTM boosting equity $R^{2}$ by up to 10\% and crypto $R^2$ by a commanding 9–12\%. The MIDAS model, as expected from its long-memory design, shows more modest improvement. 

At the 7-day horizon, long‑memory models exhibit more substantial gains. While AR and HAR still show solid improvements, the augmented LSTM achieves an accuracy boost of up to 14\%. The MIDAS model now stands out, delivering remarkable gains: its $R^{2}$ in equities surges by 11–13\%, and by as much as 23\% in cryptocurrencies, with corresponding reductions in MSE and QLIKE.

These results demonstrate the versatile power of factor augmentation, which consistently improves forecast accuracy, especially when a model's architecture aligns with the task horizon. 

\subsection{Value Across Markets: Robust in Equities, Super‑Charged in Crypto}

Factor augmentation framework proves its power and versatility across fundamentally different market structures, delivering robust gains in the diffuse equity market and a super-charged performance boost in the highly concentrated crypto market.

In the challenging equity market, where volatility drivers are highly dispersed, our framework delivers substantial value. For instance, as shown in Table \ref{tab:Equity_1d7d}, the augmented 7-day MIDAS model boosts its $R^2$ by an average of 4.7\%, while the 1-day LSTM improves its accuracy by 4.1\%, proving the framework's reliability even in such diffuse regimes. 

In cryptocurrency markets, the framework's competitive edge is even sharper. Here, where the top three factors capture nearly 95\% of market variance, the performance gains are dramatic: as detailed in Table \ref{tab:Crypto_1d7d}, the 7‑day MIDAS model's $R^2$ soars by an average of 11.1\% and the 1-day HAR model’s by 7.9\%—both more than doubling the improvements seen in equities. 

These results confirm that factor augmentation is not only a reliable enhancer in different markets, but also becomes especially powerful when risk drivers are concentrated.

\subsection{Statistically and Economically Significant Gains}

Factor augmentation delivers gains that are both substantial and statistically significant in both predictive accuracy and economic value. 
We demonstrate this on the crypto 7-day horizon task, where our framework shows the largest improvements. Three unaugmented baselines are employed: the Random Walk, an "Individual" model with asset-specific coefficients, and an "Panel" model that share coefficients across assets. 

The framework's edge in predictive accuracy is both substantial and statistically unequivocal. As shown in Table~\ref{tab: DM_R2}, while the augmented models' consistently higher $R^2$ offer strong initial evidence, the positive statistics from an MSE-based DM test provide formal proof. This confirms that the forecasting improvements are not due to random chance but are a robust gain of our methodology.

Crucially, these statistical superiority translate directly into significant economic value. As detailed in Table~\ref{tab: DM_UoW}, the augmented models consistently deliver unambiguously higher UoW outcomes, and a corresponding UoW-based DM test formally validates that these gains are statistically significant, providing clear evidence of the framework's real-world benefits for investors. 

\section{Portfolio Backtest}

To rigorously assess the framework's financial value, we conduct a volatility‑scaled pair‑trading backtest. 

\subsection{Cointegration Test}

To identify a stationary, mean-reverting spread suitable for our pairs-trading strategy, we perform a two-stage cointegration analysis on the price series within each market. 

We apply the Johansen trace test to each five-asset group. While it finds no evidence of cointegration at any rank for the equity group, the crypto sector indicates a cointegration rank of $r=2$. 
\vspace{-0.5em}
\begin{table}[H]
    \centering
    \caption{Cointegrating Vector Coefficients for Crypto}
    \begin{tabular}{l|rrrrr}
    \toprule
     & \textbf{ADA} & \textbf{BTC} & \textbf{ETH} & \textbf{LTC} & \textbf{XRP} \\
    \midrule
    CEV$_1$ & $+2.62$ & $+3.69$ & $-4.60$ & $-0.99$ & $+2.23$ \\
    CEV$_2$ & $+5.15$ & $-0.19$ & $-5.64$ & $+1.82$ & $-3.51$ \\
    \bottomrule
    \end{tabular}
    \label{tab:CEV}
\end{table} 
\vspace{-1em}

Table~\ref{tab:CEV} reports the first two cointegrating vectors ($\text{CEV}_{1,2}$), both of which load ADA and ETH with opposite signs, indicating that the dominant mean-reverting combination is the ADA–ETH spread. To further validate this pairing, we conduct pairwise Engle–Granger tests on each residual and find only ADA–ETH stationary at the 5\% level ($p = 0.045$). Consequently, our backtest focuses exclusively on the ADA–ETH spread, excluding all other pairs.

\subsection{Volatility‑Based Pairs Trading}

Our backtest framework simulates a pairs trading strategy on the cointegrated spread between ADA-ETH. The log-price spread $S_t$ is defined as: 
\vspace{-1em}
\begin{equation}
    S_t = \ln P_{\text{ADA}, t} - \beta\,\ln P_{\text{ETH}, t},
\end{equation} 
where the hedge ratio $\beta$ is estimated via rolling OLS regression. The performance is evaluated for all four benchmark models and is measured by annualized return and Sharpe ratio.

The performance is evaluated by final equity, cumulative return, and annualized Sharpe ratio.

\subsubsection{Trading Signal Generation}

Trading signals are generated from a rolling z-score of the spread, where $Z_t = \frac{S_t - \mu_{S_t}}{\sigma_{S_t}}$. We initiate a long position when $Z_t$ < -1.5 and a short position when $Z_t$ > +1.5, closing the position when the z-score reverts to zero. To limit turnover, each trade is held for at least one day. The rolling window length ($W$) is set to 70 days for the statistical models to balance responsiveness to regime changes with statistical stability, while a shorter window of 30 days is used for the LSTM to align with its input structure.

\subsubsection{Volatility-based Position Sizing} 

Our strategy employs a volatility-targeting approach, where daily position sizes are scaled to achieve a 25\% annualized volatility target. Each day, we estimate the spread volatility using our model's forecasts for the component assets' volatility \(\widehat{\sigma}_{\text{ADA},t}\), \(\widehat{\sigma}_{\text{ETH},t}\) and their return covariance:
\begin{equation}
    \widehat{\sigma}_{\text{spread},t} \;=\;
\sqrt{252} \cdot \sqrt{\widehat{\sigma}_{\text{ADA},t}^{2}
      + \beta^{2}\widehat{\sigma}_{\mathrm{ETH},t}^{2}
      - 2\beta\,\widehat{\mathrm{Cov}}(r_{\text{ADA}, t},r_{\text{ETH}, t})}.
\end{equation}

The position size is then set based on this forecast, subject to a 5:1 maximum leverage. The backtest is initialized with an equity of \$50,000 and incorporates a 5 basis point transaction cost per leg.

\subsection{Backtest Results}

To ensure a fair comparison, we evaluate statistical models on the full sample with an expanding window and the LSTM on a fixed 80:20 split. Each approach is benchmarked against a corresponding Random Walk baseline, with full results presented in Table~\ref{tab: Backtest}.

\begin{table}[!h]
  \centering
\caption{Out-of-Sample Backtest for ADA-ETH Pair}
\setlength{\tabcolsep}{6pt}
\resizebox{\linewidth}{!}{
\begin{tabular}{c  l | c c | c}
\toprule
\multicolumn{5}{c}{\textbf{Panel A: Expanding Window Models}} \\
\midrule
\textbf{Model} & \textbf{Metric} & \textbf{Unaugmented} & \textbf{Factor-Augmented} & \textbf{Improvement} \\
\midrule
\multirow{4}{*}{\shortstack{\textbf{Random Walk}\\\textbf{(Full Sample)}}} 
  & Portfolio Value (\$) & 59,161.34 & -- & -- \\
  & Ann. Return          & 5.32\%   & -- & -- \\
  & Ann. Sharpe Ratio          & 1.402     & -- & -- \\
\midrule

\multirow{4}{*}{\textbf{AR}} 
  & Portfolio Value & 58,711.22 & 60,898.40 & 3.73\% \\
  & Ann. Return         & 5.35\%   & 6.61\% & 23.55\% \\
  & Ann. Sharpe Ratio          & 1.437     & 1.443   & 0.42\% \\
\midrule

\multirow{4}{*}{\textbf{HAR}} 
  & Portfolio Value & 56,311.89 & 57,298.08 & 1.75\% \\
  & Ann. Return          & 3.86\%  & 4.52\% &  17.10\%\\
  & Ann. Sharpe Ratio          & 1.457     & 1.475   & 1.24\% \\
\midrule

\multirow{4}{*}{\shortstack{\textbf{MIDAS}\\\textbf{(k=30)}}} 
  & Portfolio Value & 58,339.32 & 58,121.55 & -0.37\% \\
  & Ann. Return          & 5.31\%   & 5.04\% &  -5.08\% \\
  & Ann. Sharpe Ratio          & 1.460     & 1.485   & 1.71\% \\
\midrule
\multicolumn{5}{c}{\textbf{Panel B: Late Sample Models}} \\
\midrule
\multirow{4}{*}{\shortstack{\textbf{Random Walk}\\\textbf{(Last 20\% Window)}}} 
  & Portfolio Value & 49,125.74 & -- & -- \\
  & Ann. Return         & -2.67\% & -- & -- \\
  & Ann. Sharpe Ratio          & -0.18   & -- & -- \\
\midrule

\multirow{4}{*}{\textbf{LSTM}} 
  & Portfolio Value & 47,427.84 & 53,306.58 & 12.40\% \\
  & Ann. Return          & -5.48\%   & 7.27\% & 232.66\% \\
  & Ann. Sharpe Ratio          & -0.467    & 0.787 & 268.52\% \\
\bottomrule
\end{tabular}%
  }
\label{tab: Backtest}

\end{table}

\subsubsection{\normalfont\bfseries Dual Gains: Amplifying Return and Sharpening Risk}

Factor augmentation decisively boosts both profitability and risk-adjusted performance across all statistical models we examine.

The short-memory models realize the most impressive gains in raw profitability. Relative to its unaugmented version, the AR specification boosts its annualized return by a remarkable 23.6\%. The HAR model shows a parallel improvement, with its return climbing by 17.1\% and its Sharpe ratio rising from 1.457 to 1.475. 

The long-memory MIDAS model, in contrast, demonstrates its value through a different channel: superior risk calibration. Although its raw return sees only modest gains—a result of its long-term design—the augmented strategy's  precise calibration of volatility drives its Sharpe ratio from 1.460 to a higher 1.485, signaling tighter risk management. 

Collectively, this evidence confirms that factor augmentation is not merely an incremental tweak but a fundamental advance in volatility forecasting. It consistently amplifies reward, sharpens risk-adjusted performance, and sets a new benchmark for the field.

\subsubsection{\normalfont\bfseries Robustness in All Market Conditions: Enhancing Gains, Reversing Losses.}

Our framework proves its value in all market conditions, systematically enhancing gains in favorable periods while acting as a powerful shield during downturns. 

During the full‐sample “bull‑market” window, when even a naïve RW was profitable, factor augmentation still delivered a distinct competitive edge. Every statistical model, when augmented, posted a higher Sharpe ratio, sharpening performance even when market conditions are already favorable.

The framework's power becomes even more pronounced in adversity. During the challenging late-sample period, when the market turned negative and the RW benchmark lost 2.7\% annually, the unaugmented LSTM suffered a -5.5\% annualized loss, likely overfitting on patterns that no longer held. 
In a stunning reversal, our factor-augmented LSTM transformed this failing strategy into a resounding success, flipping the annualized loss into a +7.3\% gain. This phenomenal turnaround propelled its Sharpe ratio from a negative -0.467 to a positive 0.787, unequivocally converting a losing proposition into a clear winner. 

This remarkable dual capability—to amplify gains in good times and decisively reverse losses in bad—undeniably proves our framework's practical utility for navigating the complexities of any market conditions.

\section{Conclusion}

This paper proposes a novel, model-agnostic Factor-Augmented Volatility Forecast framework, resolving the trade-off between computationally infeasible multivariate models, overly simplistic univariate models, and rigid static-factor models. By extracting a compact set of dynamic factors and their time-varying loadings directly from realized volatilities, our frasmework allows any baseline model to efficiently capture evolving market-wide co-movements. The framework’s robust performance—evidenced by substantial gains in forecast accuracy and superior risk-adjusted returns—combined with its computational efficiency and interpretability, establishes it as a valuable and practical solution for dynamic risk management.

\begin{acks}
Elynn Chen's research is supported in part by the NSF Award 2412577.
\end{acks}

\bibliography{ref}

\clearpage
\appendix

\section{Equity Market: Forecast Performance}

\begin{table*}[!b]
\small                            
\setlength{\tabcolsep}{4pt}       
\renewcommand{\arraystretch}{0.9} 

\caption{Equity Volatility Forecast Accuracy: 1- and 7-Day Out-of-Sample Results}
\label{tab:Equity_1d7d}
\resizebox{\linewidth}{!}{%
  \begin{tabular}{l | *{15}{c}}
    \toprule
    \multicolumn{16}{c}{\textbf{Panel A: Equity in 1‐Day Horizon}} \\
    \midrule
    \textbf{Models}
      & \multicolumn{3}{c}{\textbf{MSFT}} 
      & \multicolumn{3}{c}{\textbf{AMD}} 
      & \multicolumn{3}{c}{\textbf{INTC}} 
      & \multicolumn{3}{c}{\textbf{ORCL}} 
      & \multicolumn{3}{c}{\textbf{CSCO}} \\
    \cmidrule(lr){2-4} \cmidrule(lr){5-7} \cmidrule(lr){8-10} 
    \cmidrule(lr){11-13} \cmidrule(lr){14-16}
      & $\bm{R^2}$ & \textbf{MSE}   & \textbf{QLIKE}  
      & $\bm{R^2}$ & \textbf{MSE}   & \textbf{QLIKE}  
      & $\bm{R^2}$ & \textbf{MSE}   & \textbf{QLIKE}  
      & $\bm{R^2}$ & \textbf{MSE}   & \textbf{QLIKE}  
      & $\bm{R^2}$ & \textbf{MSE}   & \textbf{QLIKE}    \\
    \midrule
    \textbf{AR}             & 33.75 &  7.30 & 0.1756 & 32.07 &  8.79 & 0.0570 & 33.64 & 13.79 & 0.1115 & 36.49 & 23.47 & 0.1486 & 34.58 & 12.72 & 0.1133 \\
    \textbf{AR‐Aug}         & \cellcolor{GreenSupLight}34.26 &  \cellcolor{GreenSupLight}7.30 & \cellcolor{GreenSupLight}0.1741 & \cellcolor{GreenSupLight}32.33 &  \cellcolor{GreenSupLight}8.78 & \cellcolor{GreenSupLight}0.0566 & \cellcolor{GreenSupLight}34.01 & \cellcolor{GreenSupLight}13.57 & \cellcolor{GreenSupLight}0.1111 & \cellcolor{GreenSupLight}37.69 & \cellcolor{GreenSupLight}22.78 & \cellcolor{GreenSupLight}0.1443 & \cellcolor{GreenSupLight}34.78 & \cellcolor{GreenSupLight}12.55 & \cellcolor{GreenSupLight}0.1132 \\
    \textbf{HAR}            & 33.48 &  7.40 & 0.1740 & 32.87 &  8.72 & 0.0540 & 31.37 & 14.28 & 0.1139 & 37.11 & 23.37 & 0.1469 & 34.34 & 12.75 & 0.1087 \\
    \textbf{HAR‐Aug}        & \cellcolor{GreenSupLight}33.78 &  \cellcolor{GreenSupLight}7.37 & \cellcolor{yellow}0.1755 & \cellcolor{yellow}32.52 &  \cellcolor{yellow}8.84 & \cellcolor{yellow}0.0548 & \cellcolor{GreenSupLight}31.96 & \cellcolor{GreenSupLight}14.20 & \cellcolor{GreenSupLight}0.1135 & \cellcolor{GreenSupLight}37.21 & \cellcolor{GreenSupLight}23.21 & \cellcolor{GreenSupLight}0.1453 & \cellcolor{GreenMed}37.79 & \cellcolor{GreenLight}12.08 & \cellcolor{GreenSupLight}0.1049 \\
    \textbf{MIDAS (k=30)} & 35.58 &  7.16 & 0.1759 & 34.58 &  8.61 & 0.0535 & 33.89 & 14.04 & 0.1134 & 38.19 & 23.39 & 0.1504 & 36.84 & 12.53 & 0.1083 \\
    \textbf{MIDAS‐Aug}      & \cellcolor{yellow}35.44 &  \cellcolor{yellow}7.17 & \cellcolor{yellow}0.1771 & \cellcolor{yellow}34.49 &  \cellcolor{yellow}8.62 & \cellcolor{GreenSupLight}0.0533 & \cellcolor{GreenSupLight}34.29 & \cellcolor{GreenSupLight}13.95 & \cellcolor{GreenSupLight}0.1129 & \cellcolor{GreenSupLight}38.60 & \cellcolor{yellow}23.46 & \cellcolor{yellow}0.1509 & \cellcolor{GreenSupLight}36.99 & \cellcolor{GreenSupLight}12.52 & \cellcolor{yellow}0.1083 \\
    \textbf{LSTM}           & 38.87 &  4.47 & 0.1237 & 27.51 & 13.22 & 0.0577 & 42.92 & 21.12 & 0.1103 & 27.10 & 54.25 & 0.1887 & 46.74 & 18.00 & 0.0887 \\
    \textbf{LSTM‐Aug}       & \cellcolor{GreenSupLight}39.33 &  \cellcolor{GreenSupLight}4.43 & \cellcolor{GreenSupLight}0.1223 & \cellcolor{GreenSupLight}28.44 & \cellcolor{GreenSupLight}13.05 & \cellcolor{GreenSupLight}0.0564 & \cellcolor{GreenSupLight}44.16 & \cellcolor{GreenSupLight}20.73 & \cellcolor{GreenSupLight}0.1089 & \cellcolor{GreenMed}29.84 & \cellcolor{GreenSupLight}52.21 & \cellcolor{GreenSupLight}0.1876 & \cellcolor{GreenSupLight}48.08 & \cellcolor{GreenSupLight}17.79 & \cellcolor{GreenSupLight}0.0882 \\
    \midrule
    \multicolumn{16}{c}{\textbf{Panel B: Equity in 7‐Day Horizon}} \\

    \midrule
    \textbf{AR}             & 44.99 & 3.48 & 0.0946 &  47.20 & 4.12 & 0.0269 &   44.03 & 6.71 & 0.0499 & 45.67 & 12.87 & 0.0730 &   47.85 & 6.12 & 0.0514\\
    \textbf{AR‐Aug}          & \cellcolor{GreenSupLight}45.34 & \cellcolor{GreenSupLight}3.46 & \cellcolor{yellow}0.0964 & \cellcolor{GreenSupLight}47.43 & \cellcolor{GreenSupLight}4.10 & \cellcolor{GreenSupLight}0.0268 &  \cellcolor{yellow}43.87 & \cellcolor{yellow}6.73 & \cellcolor{yellow}0.0503  & \cellcolor{GreenSupLight}46.81 & \cellcolor{GreenSupLight}12.59 & \cellcolor{GreenSupLight}0.0718 & \cellcolor{GreenSupLight}49.38 & \cellcolor{GreenSupLight}5.94 & \cellcolor{GreenSupLight}0.0504 \\
    \textbf{HAR}            & 44.39 & 3.86 & 0.1084 & 42.88 & 4.48 & 0.0257 &  37.83 & 8.36 & 0.0540 & 20.99 & 21.98 & 0.1154 & 40.32 & 8.14 & 0.0558 \\
    \textbf{HAR‐Aug}         &  \cellcolor{GreenSupLight}45.71 & \cellcolor{GreenSupLight}3.81 & \cellcolor{yellow}0.1096 & \cellcolor{GreenSupLight}43.24 & \cellcolor{GreenSupLight}4.39 & \cellcolor{yellow}0.0258  & \cellcolor{GreenLight}40.07 & \cellcolor{GreenSupLight}8.06 & \cellcolor{yellow}0.0542 & \cellcolor{GreenSupLight}21.57 & \cellcolor{GreenSupLight}21.82 & \cellcolor{GreenMed}0.1036 & \cellcolor{GreenMed}45.27 & \cellcolor{GreenLight}7.47 & \cellcolor{GreenSupLight}0.0555 \\
    \textbf{MIDAS (k=30)}  &  47.26 & 3.33 & 0.0875 & 48.09 & 3.73 & 0.0254 & 45.53 & 6.21 & 0.0477  & 47.82 & 11.46 & 0.0741 & 50.02 & 5.61 & 0.0515        \\
    \textbf{MIDAS‐Aug}      & \cellcolor{GreenLight}49.61 & \cellcolor{GreenSupLight}3.18 & \cellcolor{GreenSupLight}0.0853 & \cellcolor{GreenSupLight}48.76 & \cellcolor{yellow}3.76 & \cellcolor{GreenSupLight}0.0251 & \cellcolor{GreenSupLight}46.29 & \cellcolor{GreenSupLight}6.12 & \cellcolor{GreenSupLight}0.0472 & \cellcolor{GreenSupLight}48.37 & \cellcolor{GreenSupLight}11.33 & \cellcolor{GreenSupLight}0.0711  & \cellcolor{GreenMed}55.53 & \cellcolor{GreenMed}4.99 & \cellcolor{GreenSupLight}0.0493 \\
    \textbf{MIDAS(k=50)}   &  46.35 & 3.41 & 0.0927 & 50.17 & 3.65 & 0.0251 & 45.36 & 6.27 & 0.0500 & 48.67 & 11.35 & 0.0842 & 55.30 & 5.05 & 0.0572  \\
    \textbf{MIDAS‐Aug(k=50)}    &  \cellcolor{GreenSupLight}48.38 & \cellcolor{GreenSupLight}3.28 & \cellcolor{GreenLight}0.0859 &  \cellcolor{GreenSupLight}50.49 & \cellcolor{GreenSupLight}3.63 & \cellcolor{GreenSupLight}0.0247 & \cellcolor{GreenSupLight}46.52 & \cellcolor{GreenSupLight}6.13 & \cellcolor{GreenLight}0.0472 & \cellcolor{GreenSupLight}48.70 & \cellcolor{GreenSupLight}11.35 & \cellcolor{GreenLight}0.0793  & \cellcolor{GreenSupLight}55.92 & \cellcolor{GreenSupLight}4.98 & \cellcolor{GreenSupLight}0.0545\\
    \textbf{MIDAS(k=80)}   &  49.05 & 3.15 & 0.0885  & 50.85 & 3.76 & 0.0249 & 45.13 & 6.42 & 0.0496 & 48.22 & 11.95 & 0.0750 & 50.07 & 5.72 & 0.0526 \\
    \textbf{MIDAS‐Aug(k=80)}    &  \cellcolor{GreenSupLight}51.36 & \cellcolor{GreenSupLight}3.01 & \cellcolor{GreenSupLight}0.0866  & \cellcolor{GreenSupLight}51.49 & \cellcolor{GreenSupLight}3.71 & \cellcolor{GreenSupLight}0.0239 & \cellcolor{GreenSupLight}45.68 & \cellcolor{GreenSupLight}6.35 & \cellcolor{GreenSupLight}0.0485 & \cellcolor{GreenSupLight}49.86 & \cellcolor{GreenSupLight}11.87 & \cellcolor{GreenSupLight}0.0728 & \cellcolor{GreenMed}56.50 & \cellcolor{GreenSupLight}4.98 & \cellcolor{GreenMed}0.0503  \\
    \textbf{LSTM}           & 46.38 & 3.55 & 0.1002 & 43.82 & 7.02 & 0.0337 & 42.98 & 15.13 & 0.0779 & 42.14 & 39.23 & 0.1491 & 49.47 & 13.01 & 0.0644 \\
    \textbf{LSTM‐Aug}       & \cellcolor{GreenSupLight}47.36 & \cellcolor{GreenSupLight}3.51 & \cellcolor{GreenSupLight}0.0964 & \cellcolor{GreenSupLight}44.82 & \cellcolor{GreenSupLight}6.91 & \cellcolor{GreenSupLight}0.0326 & \cellcolor{GreenLight}45.15 & \cellcolor{GreenSupLight}14.54 & \cellcolor{GreenSupLight}0.0766 & \cellcolor{GreenSupLight}42.74 & \cellcolor{GreenMed}33.29 & \cellcolor{GreenLight}0.1416 & \cellcolor{GreenSupLight}50.55 & \cellcolor{GreenSupLight}12.74 & \cellcolor{GreenSupLight}0.0628  \\
    \bottomrule
  \end{tabular}%
}
\begin{flushleft}
  \footnotesize\emph{Note:} All MSE values are reported in units of $10^{-4}$.\quad Improvement: \textcolor{GreenDark}{\rule{1.2em}{1.2em}}\ $\geq +15\%$ \quad
    \textcolor{GreenMed}{\rule{1.2em}{1.2em}}\ $\geq +10\%$\quad
    \textcolor{GreenLight}{\rule{1.2em}{1.2em}}\  $\geq +5\%$ \quad \textcolor{GreenSupLight}{\rule{1.2em}{1.2em}}\  $>\ 0\%$ \quad \textcolor{yellow}{\rule{1.2em}{1.2em}}\ $\leq -5\%$ 
\end{flushleft}

\section{Statistical Significance}

\addtocounter{table}{1}
\begin{subtable}[t]{0.48\textwidth}
    \centering
    \caption{Out-of-Sample $R^{2}$ and MSE-based DM Test for 7-day Crypto.}
    \label{tab: DM_R2}

    \setlength{\tabcolsep}{4.5pt}
    \resizebox{\linewidth}{!}{
      \begin{tabular}{c c | *{7}{>{\centering\arraybackslash}m{4em}}}
        \toprule
        \textbf{Asset} & \textbf{Model}
            & \shortstack{\textbf{Random}\\\textbf{Walk}}
            & \textbf{AR5} & \textbf{HAR}
            & \shortstack{\textbf{MIDAS}\\\textbf{(k=30)}}
            & \shortstack{\textbf{MIDAS}\\\textbf{(k=50)}}
            & \shortstack{\textbf{MIDAS}\\\textbf{(k=80)}}
            & \textbf{LSTM}\\
        \midrule
      \multicolumn{9}{c}{\textbf{Panel A: Out-of-Sample $\bm{R^2}$ (in \%)}} \\
      \midrule
      \multirow{2}{*}{\textbf{BTC}} & Individual       & 32.48 & 48.63 & 44.96 & 49.40 & 49.64 & 49.58 & 59.31 \\
                                   & Panel            & –     & 48.02 & 47.65 & 48.52 & 47.59 & 45.62 & 57.91 \\
                                   & Factor-Augmented & –     & 50.89 & 47.38 & 53.26 & 52.95 & 52.95 & 61.66 \\
      \multirow{2}{*}{\textbf{ETH}} & Individual       & 17.27 & 39.98 & 38.25 & 40.94 & 40.90 & 40.96 & 46.25 \\
                                   & Panel            & –     & 40.83 & 39.54 & 41.06 & 41.13 & 39.37 & 48.65 \\
                                   & Factor-Augmented & –     & 41.85 & 37.90 & 44.04 & 43.80 & 43.72 & 51.89 \\
      \multirow{2}{*}{\textbf{XRP}} & Individual       & 22.94 & 41.78 & 45.63 & 43.26 & 43.26 & 42.73 & 42.81 \\
                                   & Panel            & –     & 42.68 & 46.76 & 42.86 & 42.27 & 41.10 & 46.81 \\
                                   & Factor-Augmented & –     & 43.07 & 46.58 & 47.23 & 47.34 & 47.33 & 44.25 \\
      \multirow{2}{*}{\textbf{ADA}} & Individual       & 12.32 & 38.37 & 36.71 & 38.92 & 39.00 & 38.92 & 37.32 \\
                                   & Panel            & –     & 40.15 & 42.36 & 40.51 & 40.76 & 39.48 & 48.96 \\
                                   & Factor-Augmented & –     & 40.70 & 38.03 & 47.01 & 47.29 & 47.79 & 41.16 \\
      \multirow{2}{*}{\textbf{LTC}} & Individual       & 20.23 & 39.21 & 39.14 & 40.82 & 40.69 & 40.80 & 52.39 \\
                                   & Panel            & –     & 39.37 & 41.22 & 39.84 & 38.87 & 36.58 & 56.54 \\
                                   & Factor-Augmented & –     & 42.38 & 40.36 & 45.08 & 44.30 & 44.31 & 59.83 \\
      \midrule
      \multicolumn{9}{c}{\textbf{Panel B: MSE-based Diebold–Mariano Test Statistics}} \\
      \midrule
      \multirow{2}{*}{\textbf{BTC}} & Aug vs. RW & NA    & 2.46  & 1.25  & 3.34  & 3.36  & 3.32  & 3.71 \\
                                    & Aug vs. Ind  & –     & 3.13  & 1.29  & 2.18  & 2.21  & 2.06  & 4.32 \\
                                    
      \multirow{2}{*}{\textbf{ETH}}  & Aug vs. RW & NA     & 3.07  & 2.34  &                                  4.60  & 4.56  & 4.48  & 3.56 \\
                                    & Aug vs. Ind  & –     & 1.65  & 1.61  & 2.10  & 2.19  & 1.95  & 2.71 \\
                                    
    \multirow{2}{*}{\textbf{XRP}} & Aug vs. RW & NA      & 2.60  & 3.09  & 3.50  & 3.57  & 3.55  & 2.94 \\
                                    & Aug vs. Ind  & –     & 1.89  & 1.78  &  1.79 & 1.77  & 1.68  & 2.13 \\
                                    
      \multirow{2}{*}{\textbf{ADA}} & Aug vs. RW & NA      & 3.29  & 3.35  & 3.80  & 3.76  & 3.84  & 1.95 \\
                                   & Aug vs. Ind  & –     & 2.14  &  1.14 &  2.11 & 1.79  &  1.70 & 1.61 \\
                                   
      \multirow{2}{*}{\textbf{LTC}} & Aug vs. RW & NA      & 2.18  & 1.94  & 3.78  & 3.78  & 3.79  & 3.12 \\
                                   & Aug vs. Ind  & –     & 3.94  &  1.52 & 1.27  & 1.10  & 1.27  &  2.09\\
      \bottomrule
      \end{tabular}%
    }
    \begin{flushleft}
      \footnotesize\emph{Note:} Aug = Factor-Augmented,  
    Ind = Unaugmented Individual, RW = Random Walk.
    \end{flushleft}
  \end{subtable}
  \hfill
  \begin{subtable}[t]{0.48\textwidth}
    \centering
    \caption{Out-of-Sample UoW and UoW-based DM Test
             for 7-day Crypto}
    \label{tab: DM_UoW}

    \setlength{\tabcolsep}{4.5pt}
    \resizebox{\linewidth}{!}{%
      \begin{tabular}{c c | *{7}{>{\centering\arraybackslash}m{4em}}}
        \toprule
        \textbf{Asset} & \textbf{Model}
            & \shortstack{\textbf{Random}\\\textbf{Walk}}
            & \textbf{AR5} & \textbf{HAR}
            & \shortstack{\textbf{MIDAS}\\\textbf{(k=30)}}
            & \shortstack{\textbf{MIDAS}\\\textbf{(k=50)}}
            & \shortstack{\textbf{MIDAS}\\\textbf{(k=80)}}
            & \textbf{LSTM}\\
        \midrule
      \multicolumn{8}{c}{\textbf{Panel A: Out-of-Sample Utility of Wealth (in \%)}} \\
        \midrule
        BTC   & Individual & 2.66 & 3.31 & 2.87 & 3.25 & 3.23 & 3.22 & 3.66 \\
              & Panel & - & 3.32 & 3.22 & 3.27 & 3.27 & 3.27 & 3.65 \\
              & Factor-Augmented & - & 3.35 & 3.10 & 3.34 & 3.33 & 3.34 & 3.68 \\
        ETH   & Individual & 3.13 & 3.51 & 3.46 & 3.49 & 3.48 & 3.48 & 3.58 \\
              & Panel & - & 3.50 & 3.46 & 3.48 & 3.48 & 3.48 & 3.60 \\
              & Factor-Augmented & - & 3.52 & 3.47 & 3.53 & 3.52 & 3.51 & 3.65 \\
        XRP   & Individual & 2.72 & 3.14 & 3.11 & 3.08 & 3.06 & 3.04 & 2.36 \\
              & Panel & - & 3.10 & 3.13 & 3.03 & 3.03 & 3.03 & 2.48 \\
              & Factor-Augmented & - & 3.24 & 3.23 & 3.20 & 3.20 & 3.19 & 2.41 \\
        ADA   & Individual & 3.39 & 3.62 & 3.57 & 3.59 & 3.59 & 3.59 & 3.31 \\
              & Panel & - & 3.60 & 3.61 & 3.60 & 3.60 & 3.60 & 3.45 \\
              & Factor-Augmented & - & 3.65 & 3.62 & 3.62 & 3.62 & 3.61 & 3.35 \\
        LTC   & Individual & 3.23 & 3.59 & 3.52 & 3.56 & 3.55 & 3.55 & 3.68 \\
              & Panel & - & 3.56 & 3.56 & 3.56 & 3.56 & 3.56 & 3.69 \\
              & Factor-Augmented & - & 3.60 & 3.57 & 3.59 & 3.58 & 3.58 & 3.71 \\
        \midrule
        \multicolumn{8}{c}{\textbf{Panel B: UoW-based Diebold-Mariano Test Statistics}} \\
        \midrule
          \multirow{2}{*}{\textbf{BTC}} & Aug vs. RW & NA    & 3.35 & 3.10 & 3.34 & 3.33 & 3.34 & 3.68 \\
                                        & Aug vs. Ind  & –     & 1.46  & 1.43  & 1.96  & 2.17  & 2.06  & 2.93 \\
                                        
          \multirow{2}{*}{\textbf{ETH}}  & Aug vs. RW & NA     & 3.52 & 3.47 & 3.53 & 3.52 & 3.51 & 3.65 \\
                                        & Aug vs. Ind  & –     & 1.55  & 0.668  & 1.84  & 1.79  & 1.74  & 2.38 \\
                                        
        \multirow{2}{*}{\textbf{XRP}} & Aug vs. RW& NA      & 3.24 & 3.23 & 3.20 & 3.20 & 3.19 & 2.41 \\
                                        & Aug vs. Ind & –     & 1.82  & 1.20  &  1.52 & 1.62  & 1.70  & 1.63\\
                                        
          \multirow{2}{*}{\textbf{ADA}} & Aug vs. RW & NA   & 3.65 & 3.62 & 3.62 & 3.62 & 3.61 & 3.35 \\
                                       & Aug vs. Ind  & –     & 2.43  &  1.13 &  2.17 & 2.30  &  2.24 & 2.32 \\
                                       
          \multirow{2}{*}{\textbf{LTC}} & Aug vs. RW & NA      & 3.60 & 3.57 & 3.59 & 3.58 & 3.58 & 3.71 \\
                                       & Aug vs. Ind  & –     & 1.67  &  1.72 & 1.89  & 2.24 & 1.89  &  2.44\\
          \bottomrule
      \end{tabular}%
    }
    \begin{flushleft}
      \footnotesize\emph{Note:} Definitions same as table (a).
        \end{flushleft}
    \end{subtable}
\end{table*}

\end{document}